%% file: adk_lh.tex
\documentclass[%
 aps, pra,
 amsmath,amssymb,
 reprint,%
]{revtex4-1}

\usepackage{graphicx}
\usepackage{dcolumn}
\usepackage{bm}

\usepackage[utf8]{inputenc}
\usepackage[T1]{fontenc}
\usepackage{mathptmx}

\begin{document}


\preprint{AIP/123-QED}

\title[The Ultrafast Kerr Effect in Anisotropic and Dispersive Media]{The Ultrafast Kerr Effect in Anisotropic and Dispersive Media}

\author{Lucas Huber}
\email[To whom correspondence should be addressed: ]{lh3019@columbia.edu}
\affiliation{Department of Chemistry, Columbia University, New York, NY-10027, USA}
\author{Sebastian Maehrlein}
\affiliation{Department of Chemistry, Columbia University, New York, NY-10027, USA}
\author{Feifan Wang}
\affiliation{Department of Chemistry, Columbia University, New York, NY-10027, USA}
\author{Yufeng Liu}
\affiliation{Department of Chemistry, Columbia University, New York, NY-10027, USA}
\author{X.-Y.\ Zhu}
\affiliation{Department of Chemistry, Columbia University, New York, NY-10027, USA}

\date{\today}

\begin{abstract} 
The ultrafast optical Kerr effect (OKE) is widely used to investigate the structural dynamics and interactions of liquids, solutions and solids by observing their intrinsic nonlinear temporal responses through nearly-collinear four-wave mixing (FWM). Non-degenerate mixing schemes allow for background free detection and can provide information on the interplay between a material's internal degrees of freedom. 
Here we show a source of temporal dynamics in the OKE signal that is not reflective of the intrinsic nonlinear response but arises from
group index and momentum mismatch. It is observed in two-color experiments on condensed media with sizable spectral dispersion, a common property near an optical resonance.
In particular birefringence in crystalline solids is able to entirely change the character of the OKE signal via the off-diagonal tensor elements of the nonlinear susceptibility. 
We develop a detailed description of the phase-mismatched ultrafast OKE and show how to extract quantitative information on the spectrally resolved birefringence and group index from time-resolved experiments in one and two dimensions. 
\end{abstract}

\maketitle

\section{Introduction}

The optical Kerr effect is commonly described as intensity dependent modulation of the real part of the refractive index 
$\Delta n = n_2 I = 3 \chi^{(3)}/(4 n^2_0 \epsilon_0 c_0) I$, where $n_0$ is the refractive index, $\epsilon_0$ is the vacuum permittivity and $c_0$ is the speed of light \cite{boyd2008}. It can be considered as a special case of the more general phenomenon of FWM via the 3rd order nonlinear susceptibility $\chi^{(3)}$. Its isotropic manifestation gives rise to the well known effects of self-phase modulation and Kerr-lensing which is the basis of the most common passively mode-locked laser systems \cite{Brabec92}. 
Among all nonlinear processes, FWM based on the interaction of electric fields is the most abundant as the associated polar tensor of the 4th rank $\chi^{(3)}_{ijkl}$ 
is symmetry allowed in all 32 crystallographic point groups \cite{armstrong62, birss}. 
In anisotropic materials, 
the Kerr effect becomes a tensorial problem requiring to discard 
scalar terms like $n_2$ and intensity $I$ and instead use a general FWM formalism. Here, we assume the fully collinear case in which $\chi^{(3)}$ gives rise to a time-dependent nonlinear polarization inside a material at location $z$ according to
\begin{align}
\begin{split}
P^{(NL)}_i (t,z) = &\epsilon_0\int\limits^t_{-\infty}\mathrm{d}t'\int\limits^{t'}_{-\infty}\mathrm{d}t''\int\limits^{t''}_{-\infty}\mathrm{d}t''' \chi^{(3)}_{ijkl}(t,t',t'', t''',z) \\
&\times E_j(t',z)E_k(t'',z)E_l(t''',z).
\label{eq1}
\end{split}
\end{align}
The sum convention is used, summing over all possible polarization combinations of the three independent complex light fields $E_{j}$, $E_{k}$, $E_{l}$ and the non-vanishing third order tensor components of $\chi^{(3)}_{ijkl}$. 
Each tensor component may have a specific temporal evolution with respect to the three contributing fields 
depending on $t', t'', t'''$ while the result will evolve in time $t$. Studying the corresponding hyperpolarizability and its temporal evolution provides intricate details on the interplay of various degrees of freedom in a condensed medium and launched the field of multi-dimensional optical spectroscopy \cite{Tokmakoff1997, Mukamel2000, Mukamel1995}. In contrast, conventional Kerr spectroscopy is based on a one-dimensional pump-probe scheme (1D), where two fields in Eq.\ \ref{eq1} are provided by the same pump-pulse. It is commonly used to study the coupling of molecules to their environment in disordered systems such as solutions and plastic crystals \cite{McMorrow1988, Righini993}. In the solid state, the experimental technique is equivalent to time-resolved stimulated Raman spectroscopy of anisotropic modes \cite{merlin1997, Yan1987a, Yan1987b} and recently there was interest in applying the method to study phonon dynamics in lead-halide perovskites \cite{Zhu1409, Miyata2017}. 
There are, however, several caveats when applying Kerr spectroscopy to dispersive and anisotropic solids that are not of concern in liquids of low optical density. 
Light propagation effects in FWM and their influence on lifetime or line shape measurements have previously been discussed in the context of two-level quantum systems \cite{Saikan1987, Kinrot1994, Kinrot1995}. Phase-matching considerations were also reported in specific multi-level systems \cite{Pantke1993, Schillak1993,Bakker1994, Schulze1995, Sunkgkyu1999, Thompson2000, Yetzbacher2007, Cho2009} and recently in non-degenerate triple sum-frequency mixing \cite{Morrow2017}. 
However, to our knowledge, the role of anisotropy in ultrafast Kerr spectroscopy has not been considered in detail.\\
This article is therefore going to discuss the significant role of the 
spatial dependence of $P^{(NL)}_i(t,z)$ in time-resolved studies in a collinear geometry which is typically framed as a \textit{phase-matching} problem \cite{Maker1962}. We will show how a local nonlinear polarization is subject to spectral dispersion and how the time-resolved signal entirely changes character in the presence of anisotropy. Furthermore, we will show how to experimentally resolve the spectral dependence of the phase-mismatch $\Delta k$ by expanding conventional Kerr spectroscopy to a two-dimensional scheme (2D) \cite{Maehrlein2020}. 
Emphasizing the generality of this effect, we will discuss this phenomenon using a classical description that is solely based on a material's dispersive refractive index tensor and can hence be applied straightforward to arbitrary material systems. Lastly, we propose and describe in detail how to use the information contained in these spectra for birefringence spectroscopy. \\

\section{Theory}

\subsection{The anisotropic \& dispersive Kerr effect}
\label{sec:adk}
In this section, we will derive expressions that allow to calculate the Kerr effect in experimental configurations with one or two temporal dimensions.
While Eqn.\ \ref{eq1} allows for a range of nonlinear processes to occur, the observation of specific pathways is defined by the experimental setup. 
In a pump-probe geometry, 
a strong excitation pulse with a spectrum centered at $\omega_1$ is typically 
combined with a weaker probe pulse at $\omega_3$, contributing as $E^{(1)}_iE^{(2)}_j$ and $E^{(3)}_k$ to Eqn.\ \ref{eq1}, respectively. 
In the following we will discuss ultrafast two-color Kerr spectroscopy in which excitation and probe beams 
are delayed by $t_{pr}$ with respect to each other and are combined inside a material 
to result in a nonlinear field. 
Only the spectral range of the probe needs to enter the detection and non-degenerate pump-fields can be filtered out allowing for fully collinear geometries. 

\begin{figure*}[bt]
	\centering
		\includegraphics[width=1.0\textwidth]{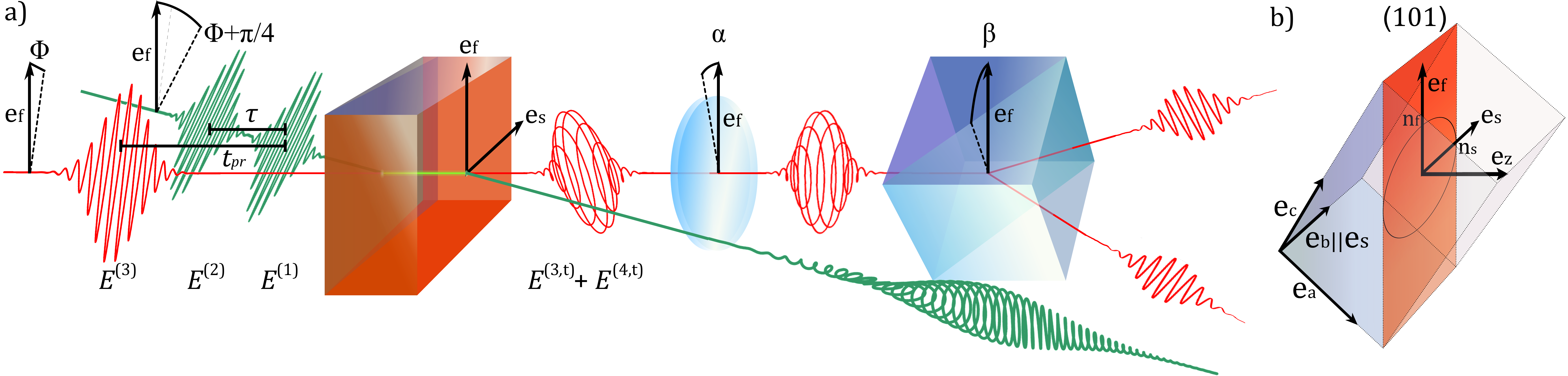}
	\caption{a) Schematic representation of an ultrafast OKE experiment in transmission geometry with a balanced detection. The assumed pump (green) -probe (red) geometry is collinear, but depicted at an angle for clarity. All fields and rotations are defined with respect to the fast and slow material axes $\mathbf{e}_{\mathrm{f}}$ and $\mathbf{e}_{\mathrm{s}}$ for propagation along $\mathbf{e}_{\mathrm{z}}$.
	At energies close to the band-gap, initially compressed excitation pulses experience strong dispersion. Birefringent crystals (orange) act as arbitrary wave-retarders even in the absence of excitation. A $\lambda/2$-waveplate (light-blue) is set to an angle $\alpha$ to balance the unperturbed transmitted probe intensity detected after a Wollaston-type polarizer at an angle $\beta = \phi - \pi/4$. b) The orientation of the exemplary orthorhombic CsPbBr$_3$ single crystal unit-cell with respect to the experimental geometry. The crystal surface is given by the reciprocal lattice vector (101).}
	\label{figdetection}
\end{figure*}
Such a 1D experiment can be expanded to obtain intra-pulse spectral resolution in the excitation fields by employing a double pump scheme with an additional delay $\tau$ in analogy to 2D electronic state spectroscopy as described for the specific case discussed here by Maehrlein et al.\ \cite{Maehrlein2020} and more generally by Shim and Zanni \cite{Zanni2009}. 
By keeping only contributions to the nonlinear polarization that oscillate at frequencies close to $\omega_3$, Eqn.\ \ref{eq1} reduces to
\begin{widetext}
\begin{align}
\begin{split}
P^{(NL)}_i (t,z) = \epsilon_0\int\limits^t_{-\infty}\mathrm{d}t'\int\limits^{t'}_{-\infty}\mathrm{d}t''\int\limits^{t''}_{-\infty}\mathrm{d}t''' \chi^{(3)}_{ijkl}(t,t',t'', t''',z) 
 \left[E^{(1)}_{j}(t',z)E^{(2)*}_{k}(t'' - \tau,z) + E^{(1)*}_{j}(t',z)E^{(2)}_{k}(t'' - \tau,z)\right] 
 E^{(3)}_{l}(t''' - t_{pr},z),
\label{eqNLP}
\end{split}
\end{align}
\end{widetext}
where the nonlinear polarization is induced by the product of two complex pump field components $E^{(1)}_{j}E^{(2)*}_{k}$ (and their conjugate)
and the probe field $E^{(3)}_{l}$. Note that anisotropy will lift the degeneracy of the two conjugate terms when $j\ne k$, giving rise to two distinct contributions. For $\tau= 0$ s the expression describes the case of a 1D pump-probe experiment.\\

For simplicity we now consider a collinear transmission geometry experiment where the beams are loosely focused through a homogeneous but anisotropic sample of intermediate thickness and for which the plane wave approximation is valid. Furthermore, we consider a planar sample at normal incidence.
In this geometry, the polarization treatment in an isotropic, uni- or biaxial material is solely defined by the fast 
(minor) and slow 
(major) axes of the refractive index ellipse corresponding to the direction of propagation $\bm{k}$ 
with respect to the crystal axes, see Fig.\ \ref{figdetection} b). We can then reduce the polarization description to the two dimensional Jones formalism with base vectors $\mathbf{e}_\mathrm{f}$ and $\mathbf{e}_\mathrm{s}$ aligned along the fast and slow axes, respectively.  


Reproducing commonly used experimental conditions, we further assume pump and probe fields to be linearly polarized at an angle of 45$^\circ$ with respect to each other before entering the sample, as depicted in Fig.\ \ref{figdetection} a). 
In the decoupled low field limit, the three linear fields inside the sample can be calculated at any location $z$ and time $t$ 
\begin{align}
\begin{split}
E^{(1)}_{i}(t,z) = \int\limits_{0}^{\infty} &\mathrm{t}_i(\omega)A^{(1)}_{i}(\omega)e^{-\mathrm{i}[\omega t - k_i(\omega) z]} \\&\times\left[1 +R_i(\omega,z) \right] \mathrm{d}\omega, \\
E^{(2)}_{i}(t, \tau,z) = \int\limits_{0}^{\infty} &\mathrm{t}_i(\omega)A^{(2)}_{i}(\omega)e^{-\mathrm{i}[\omega (t-\tau) - k_i(\omega) z]} \\ &\times\left[1 +R_i(\omega,z) \right] \mathrm{d}\omega, \\
E^{(3)}_{i}(t, t_{pr},z) = \int\limits_{0}^{\infty} &\mathrm{t}_i(\omega)A^{(3)}_{i}(\omega)e^{-\mathrm{i}[\omega (t - t_{pr}) - k_i(\omega) z]} \\ &\times\left[1 +R_i(\omega,z) \right] \mathrm{d}\omega, \\
R_i(\omega,z) = \mathrm{r}_i(\omega) [1 + &e^{2 \mathrm{i} z k_i(\omega)}]\frac{e^{2 \mathrm{i} [d-z]k_i(\omega)} }{1 - \mathrm{r}^2_i(\omega) e^{2 \mathrm{i} d k_i(\omega)}},
\end{split}
\label{eqfields}
\end{align}
where $A^{(n)}_i(\omega)$ is the spectral amplitude of field $n$ in the coordinate system described above with $i \in [\mathrm{f}, \mathrm{s}]$ and $\mathrm{t}_i(\omega)$ and $\mathrm{r}_i(\omega)$ denote the transmission and reflection coefficients, respectively. The term $R_i(\omega,z)$ accounts for all orders of internal reflections, with the explicit solution given for a planar sample of thickness $d$. 

The relevant quantity in this context is the wave vector $k_i(\omega) = n_i(\omega)\omega/c_0$. Optical anisotropy implies birefringence $\Delta n = n_\mathrm{s}(\omega) - n_\mathrm{f}(\omega) \ne 0$ that causes a phase retardation between $\mathbf{e}_\mathrm{f}$- and $\mathbf{e}_\mathrm{s}$-polarized fields that evolves inside the material and will be imprinted onto the local nonlinear polarization in particular via the off-diagonal tensor elements of $\chi^{(3)}_{ijkl}$ according to Eqn.\ \ref{eqNLP}. Unless they are ideally polarized along $\mathbf{e}_\mathrm{f}$ or $\mathbf{e}_\mathrm{s}$, both pump and probe beams are subject to this effect. 

We will now consider the illustrative case where the nonlinear response 
is homogeneous throughout the sample volume and is entirely defined by a real-valued non-resonant response with no inner temporal structure (for a more realistic implementation of a quasi-instantaneous electronic response in direct-bandgap semiconductors, the universal $\chi^{(3)}$-dispersion derived by Sheik-Bahae et al.\ can be employed \cite{Sheik1990, Sheik1991}). In this case the system does not possess Raman-modes, electronic resonances or alike and is simply given by the instantaneous hyperpolarizability $\chi^{(3)}_{ijkl}(t, t', t'', t''', z) = \chi^{(3)}_{ijkl} \delta(t-t')\delta(t'-t'')\delta(t'' - t''')$. Eqn.\ \ref{eqNLP} then simplifies to the sum of products
\begin{align}
\begin{split}
P^{(NL)}_i(t,t_{pr},\tau, z) = \epsilon_0\chi^{(3)}_{ijkl} [&E^{(1)}_{j}(t,z)E^{(2)*}_{k}( t-\tau,z)\\
 + &E^{(1)*}_{j}(t,z)E^{(2)}_{k}( t-\tau,z)] \\
\times &E^{(3)}_{l}(t-t_{pr},z).
\end{split}
\label{eqnonlinearpol}
\end{align}
To facilitate the correct treatment of the phase evolution, $P^{(NL)}$ can be transformed into the frequency domain
\begin{align}
P^{(NL)}_i(\omega, t_{pr}, \tau, z) = \int_{-\infty}^{\infty}{P^{(NL)}_i(t, t_{pr}, \tau, z)e^{\mathrm{i}\omega t}\mathrm{d}t}.
\label{eqft1}
\end{align}
The local nonlinear polarization acts as a source of a new field following the one-dimensional inhomogeneous wave-equation 
\begin{align}
\left[\nabla^2 + k_i^2(\omega)\right] E^{(4)}_i(\omega, t_{pr}, \tau, z) = -\mu_0\omega^2 P^{(NL)}_i(\omega, t_{pr}, \tau, z),
\label{eqiwe}
\end{align}
where $\mu_0$ is the vacuum permeability. In the weak interaction limit, the field $E^{(4)}$ does not further contribute to the nonlinear polarization and the wave-equations of $E^{(1)}$, $E^{(2)}$, $E^{(3)}$ and $E^{(4)}$ are decoupled, which was already used above in Eqn.\ \ref{eqfields}. 
Considering only the field emission in forward direction 
we can write 
\begin{align}
E^{(4)}_i(\omega, t_{pr}, \tau, z) = A^{(4)}_i(\omega, t_{pr}, \tau,z) e^{\mathrm{i} k_i(\omega) z}.
\label{eqnlfield4}
\end{align}
In the slowly varying amplitude approximation the local contribution from the nonlinear polarization can be calculated as \cite{boyd2008}
\begin{align}
\frac{\mathrm{d} A^{(4)}_i}{\mathrm{d} z}(\omega, t_{pr}, \tau, z) = \frac{\mathrm{i} \omega^2 e^{-\mathrm{i} k_i(\omega) z}}{2 k_i(\omega) \epsilon_0 c_0^2} P^{(NL)}_i(\omega, t_{pr}, \tau, z).
\label{eqsvaa}
\end{align}
The full emitted nonlinear field amplitude is then obtained by integration 
\begin{align}
\begin{split}
A^{(4,\mathrm{t})}_i(\omega,t_{pr}, \tau,d) = \mathrm{t}_i(\omega)\int^{d}_0 \frac{\mathrm{d} A^{(4)}_i}{\mathrm{d} z}(\omega, t_{pr}, \tau, z) \mathrm{d}z.
\end{split}
\label{eqFT}
\end{align}
Eqn.\ \ref{eqFT} spatially integrates over the phase-mismatch between induced polarization and the field at $\omega$. The result is thus expected to be a slowly oscillating function of thickness $d$ and the delays $t_{pr}$ and $\tau$. This is in close similarity to the spatial phase-matching fringes observed in static second-harmonic generation (SHG) by Maker and Terhune \cite{Maker1962} and in the acoustically perturbed time-dependent SHG by Huber et al.\ \cite{lhuber2015}. \\
A heterodyne detection scheme, such as balanced detection, then allows to observe the emitted field via interference with the co-propagating probe field. The implementation of the balanced detection in the calculation is described in App.\ \ref{app:balanced}. Under balanced conditions and assuming small nonlinear field amplitudes, the detected non-equilibrium signal is
\begin{align}
\begin{split}
S \propto \int\operatorname{Re}[&(\bm{J}_1 \mathbf{E}^{(3,\mathrm{t})}) \cdot (\bm{J}_1\mathbf{E}^{(4,\mathrm{t})*}) \\
 &- (\bm{J}_2 \mathbf{E}^{(3,\mathrm{t})}) \cdot (\bm{J}_2\mathbf{E}^{(4,\mathrm{t})*})]\mathrm{d}\omega, 
\end{split}
\label{eqsignal}
\end{align}
where $\bm{J}_{1,2}$ are Jones matrices that account for the combined action of the polarization optics for the two detection channels. 
The method described above allows to calculate the time-resolved Kerr signal in materials of arbitrary dispersion and thickness in the weak field limit. 
%
%
\section{Results}
\subsection{Comparison to Experiments}
\begin{figure}[tb]
	\centering
		\includegraphics[width=0.485\textwidth]{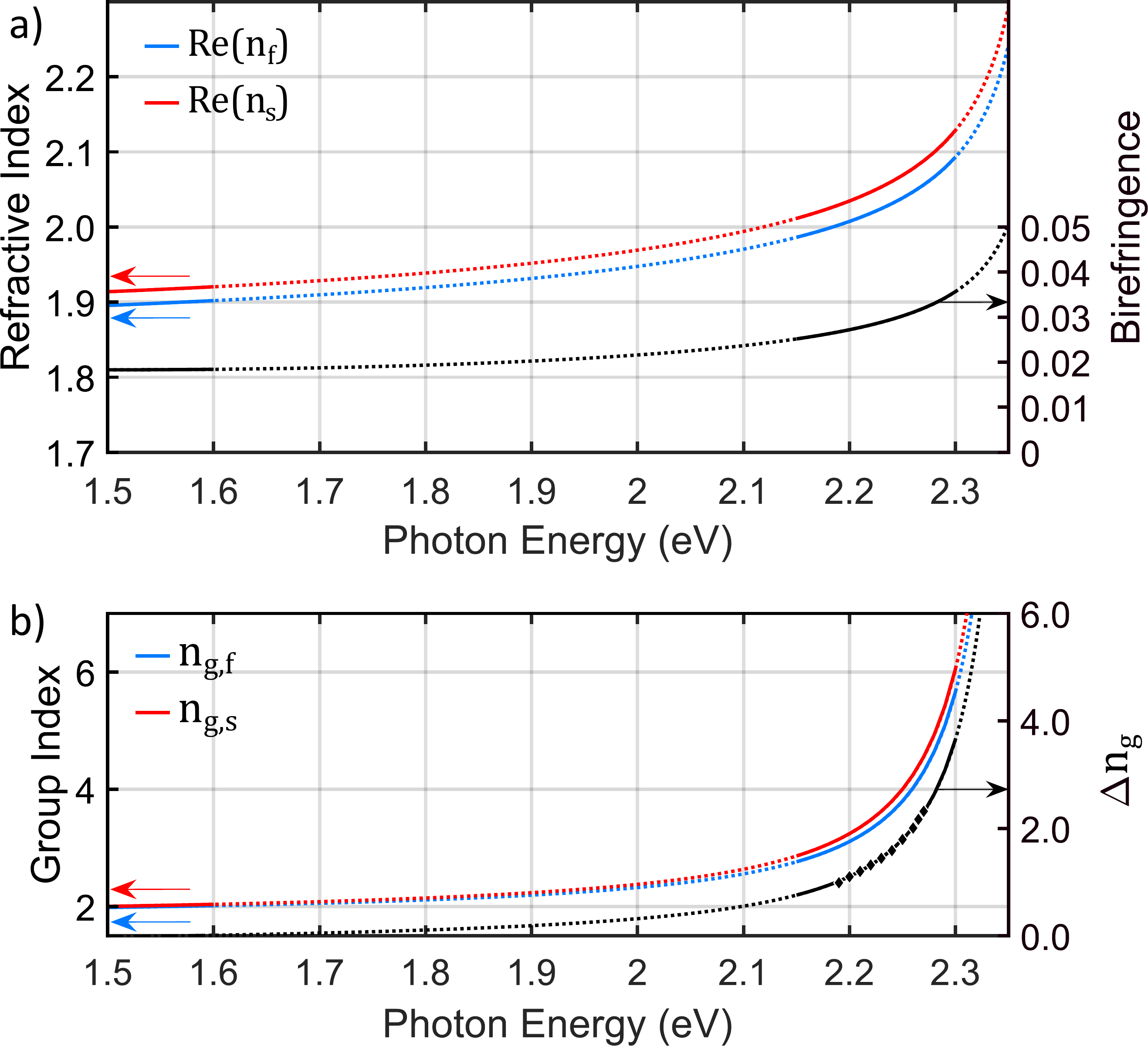}
	\caption{a) The real part of the refractive index of CsPbBr$_3$ at room temperature along fast and slow crystal axes. The function is the result of the fit to experimental data described below and in App.\ \ref{app:index}. The spectral range relevant for the fit is indicated as solid lines. The corresponding birefringence is shown in black on the right ordinate. b) The group indices corresponding to the refractive indices are shown in a). Strong dispersion near the band-gap leads to pump pulse propagation several times slower than at probe frequencies. The group indices also show significant anisotropy. Black diamonds indicate experimentally determined values of $\Delta n_{\mathrm{g}} = n^{(1)}_{\mathrm{g}}(\omega_1) - n^{(3)}_{\mathrm{g}}(\omega_3)$ based on the reflection times $t_1$.}
	\label{figrefindex}
\end{figure}

As an example, we discuss the Kerr signal from single crystal lead halide perovskite CsPbBr$_3$ in its orthorhombic phase at room temperature. 
The 2D Kerr response of an anisotropic material with quasi-instantaneous hyperpolarizability can be calculated for varying delays $t_{pr}$ and $\tau$ based on the set of equations \ref{eqfields} through \ref{eqsignal}.
The spectrally dependent refractive indices $n_{\mathrm{f},\mathrm{s}}$ employed for the calculation are given in Fig.\ \ref{figrefindex} a). The group index in Fig.\ \ref{figrefindex} b) is derived from the refractive index according to $n_{\mathrm{g},i}(\omega) = n_i(\omega) + \omega \partial n_i(\omega)/\partial\omega$.
In order to compare with our experimental data, a crystal thickness of 480 $\mu$m is assumed for the calculation. The crystal's point group symmetry defines which tensor components of $\chi^{(3)}$ can contribute to the signal. Orthorhombic CsPbBr$_3$ has space group $Pnma$ with point group $mmm$ \cite{RakitaCsPbBr3}. Without invoking further symmetries of the excitation fields, this allows for 21 independent tensor components \cite{birss}. In the absence of resonances as described by Eq.\ \ref{eqnonlinearpol}, we may allow for arbitrary permutations of the excitation fields, reducing the number of independent coefficients to 9. We compare the calculation to experiments on a (101) facet, where a specific projection of the intrinsic tensor elements is observed as described in App.\ \ref{apptensor}.

For simplicity, the calculation assumes all allowed elements of $\chi^{(3)}$ to be the same constant. Note that this implies that anisotropy is only considered in the linear refractive index. 
With this, all relevant material parameters are specified. \\
In close similarity to experimental conditions, the simulation uses two identical initially transform-limited Gaussian pulses centered at 2.237 eV with a FWHM bandwidth of 107 meV as excitation pulses $E^{(1)}$ and $E^{(2)}$ and a probe pulse $E^{(3)}$ centered at 1.57 eV with a 44 meV bandwidth. The geometric definitions for the linearly polarized pulses are given in Fig.\ \ref{figdetection} a). The angle between initial probe polarization and fast crystal axis is set to $\phi=0^\circ$. In the experiment this angle was small but different from zero. Experiment and simulation used the "backward" scanning direction for the excitation delay $\tau$, where for an increasing $\tau$ the delay between scanned pump and probe pulse increases.  
Furthermore, to mimic the experimental implementation we employ a standard 4-sequence phase-cycling scheme
\begin{align}
\begin{split}
\tilde{S} = &\left(S(E_1, E_2) - S(E_1, -E_2)\right) \\&+ \left(S(-E_1, -E_2)-S(-E_1, E_2)\right), 
\end{split}
\label{eqphasecycl}
\end{align}
where instead of adding a phase delay as done experimentally \cite{Zanni2009, Hamm2011}, the simulation fields are simply inverted. 
To improve numerical efficiency, we employ rotating frames by adding phase increments to the phase evolution of the delayed excitation fields $\Delta \phi_{\textrm{RF1}} = -\omega_{\textrm{RF1}} \tau$, as well as for the real time $\Delta \phi_{\textrm{RF2}} = -\omega_{\textrm{RF2}} t$ in all fields of Eqn.\ \ref{eqfields} \cite{Zanni2009}. While the former is also implemented experimentally to allow for reduced sampling of parameter $\tau$, the experiment naturally evolves in real time $t$. The frequencies were chosen to be $\omega_{\rm{RF1}} = 2 \pi \times 500 $ THz and $\omega_{\rm{RF2}} = 2 \pi \times 350 $ THz in order to lie below the lowest relevant frequencies of the excitation and probe fields, respectively. 
The information contained in the response to the two-pulse excitation as function of $\tau$ allows us to assign the nonlinear response to the excitation spectrum by Fourier transformation, in complete analogy to the experiment \cite{Maehrlein2020}. \\

\begin{figure*}[tbh]
	\centering
	 \includegraphics[width=1.00\textwidth]{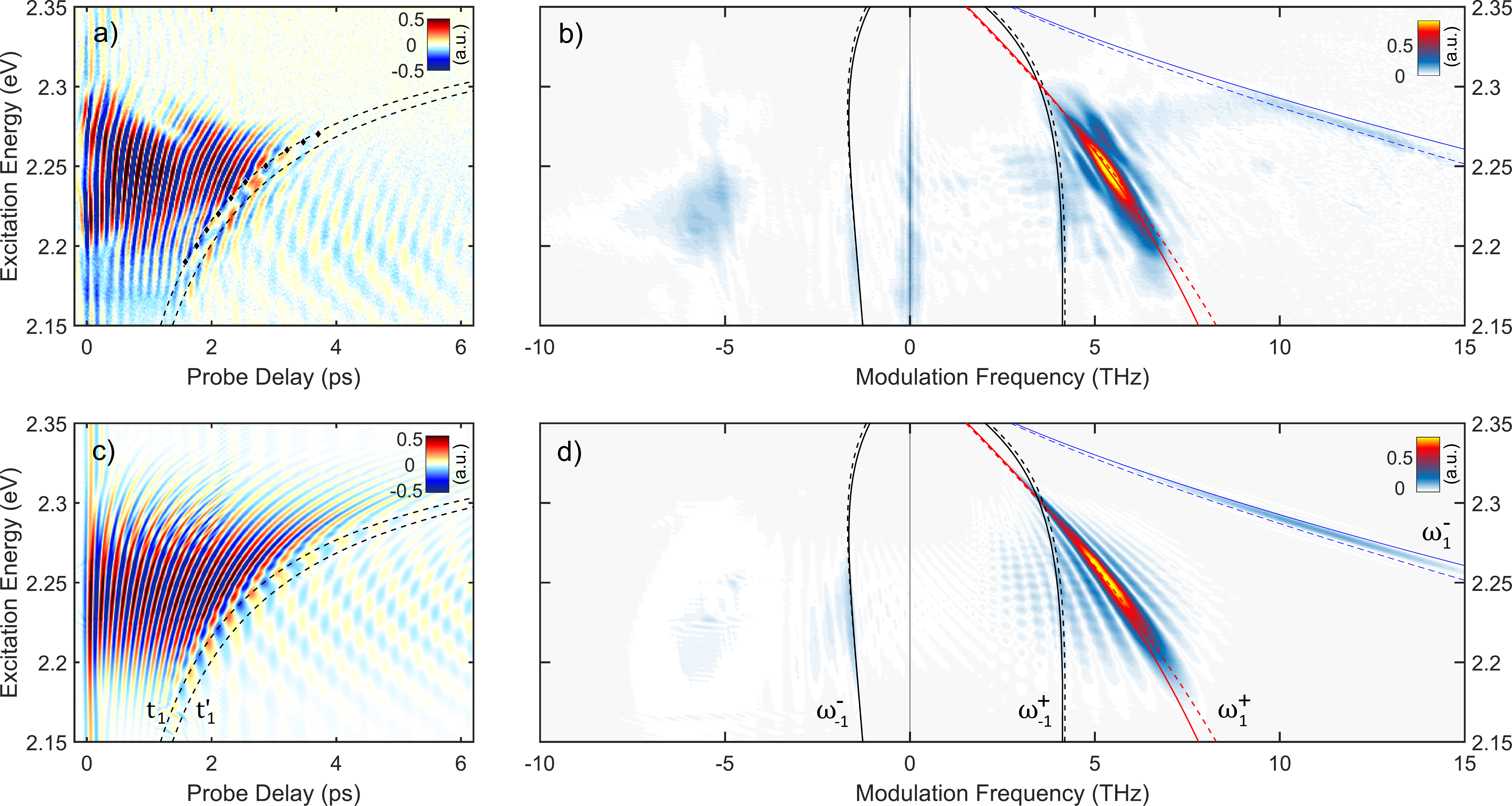}
\caption{a) Experimental results of the time and excitation energy-resolved Kerr effect obtained on a $480$ $\mu$m thick CsPbBr$_3$ sample at room temperature. Diamonds indicate the experimentally determined timings of the first internal reflection of the excitation fields \cite{Maehrlein2020}.  
	b) Corresponding 2D frequency domain showing two strongly dispersive bands centered around 6 and 11 THz as well as two less dispersive bands around -1.6 and 4.1 THz. 
	c), d) Simulated energy-time domain and 2D frequency response based on the set of equations \ref{eqfields} through \ref{eqsignal} for a $d = 480$ $\mu$m thick CsPbBr$_3$ sample at $\phi = 0^\circ$. A birefringent refractive index model was fitted to the experimental data in a) and b) according to Eqn.\ \ref{app:fappr3} and \ref{app:t1}.	For comparison, the fitted results for $t_1$ and $t'_1$ are shown as dashed lines in a) and c). The frequencies $\omega^{\pm}_\mathrm{q}$ are shown in b) and d) for Eqn.\ \ref{app:fappr3} and its approximation \ref{app:fappr4} in solid and dashed lines, respectively. The energy assignment of all lines was corrected according to the sum in Eqn.\ \ref{eq:enass} corresponding to the "backward" scanning direction of $\tau$, see Sec.\ \ref{sec:enass}. }
\label{figresults}
\end{figure*}

Numerical and experimental results are compared in Fig.\ \ref{figresults} a) and b), respectively, and show a striking similarity in time as well as frequency domain behavior. Most remarkably, the time dependent signal shows two temporal regions, $t<t_1$ and $t> t'_1$, with different oscillatory behavior in the THz range separated by a short transition period $t_1 < t \le t'_1$. Both oscillation frequencies and times at which these transitions occur show a strong excitation energy dependence.
Furthermore, the transition is abrupt leading to truncation bands next to the main feature in the frequency domain. Prominent spectral features are the strongly dispersive bands around 6 THz and 11 THz, that can be assigned to region $t<t_1$, and the weakly dispersive bands at lower frequencies near -1.6 THz and 4.1 THz originating in $t>t'_1$. 
The zero-frequency feature in the experiment is not found in the calculation for $\phi = 0$ but can be reproduced for small deviations from ideal polarization alignment.

\section{Discussion}

\subsection{Analytic derivation of modulation frequencies}

To gain a better understanding of how anisotropic field propagation is shaping the nonlinear Kerr response, we now derive approximate analytic expressions for the frequencies observed in a 1D pump-probe experiment.    
Expanding on the derivations for ultrafast two and three wave sum-frequency generation by Morrow et al.\ \cite{Morrow2017} and Angerer et al.\ \cite{Angerer1999}, we derive expressions for the frequencies observed in the ultrafast Kerr response, starting from a first-order Taylor expansion of the wave vectors in the frequency domain
\begin{align}
\begin{split}
k^{(1)}_j(\omega') &\approx n_j^{(1)} \omega_1/c_0 + n_{\mathrm{g},j}^{(1)}[\omega' - \omega_1]/c_0, \\
k^{(2)}_k(\omega'') &\approx n_k^{(2)} \omega_2/c_0 + n_{\mathrm{g},k}^{(2)}[\omega'' - \omega_2]/c_0, \\
k^{(3)}_l(\omega''') &\approx n_l^{(3)} \omega_3/c_0 + n_{\mathrm{g},l}^{(3)}[\omega''' - \omega_3]/c_0, \\
k^{(4)}_i(\omega) &\approx n_i^{(4)} \omega_4/c_0 + n_{\mathrm{g},i}^{(4)}[\omega - \omega_4]/c_0.
\label{app:wave_vector}
\end{split}
\end{align}
where $n^{(h)}_i$ refers to the refractive index at frequency $\omega_h$ along polarization direction $i$. 
By only considering the first order of the expansion, we now neglect group velocity dispersion and higher orders.
To further simplify the calculation we assume Gaussian light pulses of identical bandwidths $\sigma$
\begin{align}
\begin{split}
E^{(1)}_j(\omega') &= A^{(1)}_j e^{ \mathrm{i} \mathrm{q} k^{(1)}_j(\omega') z} e^{-(\omega' - \omega_1)^2/(2 \sigma^2)}, \\
E^{(2)}_k(\omega'') &= A^{(2)}_k e^{ \mathrm{i} \mathrm{q} k^{(2)}_k(\omega'') z} e^{-(\omega'' - \omega_2)^2/(2 \sigma^2)}, \\
E^{(3)}_l(\omega''',t_{pr}) &= A^{(3)}_l e^{\mathrm{i} k^{(3)}_l(\omega''') z} e^{-(\omega''' - \omega_3)^2/(2 \sigma^2)} e^{-\mathrm{i} \omega''' t_{pr}}, 
\label{app:gauss_fields}
\end{split}
\end{align}
where the delay of the probe field $E^{(3)}$ is accounted for by the phase $\omega''' t_{pr}$. The parameter $\mathrm{q}$ is introduced to allow for co- and counter-propagating excitation fields with respect to the probe with $\mathrm{q}=1$ and $\mathrm{q}=-1$, respectively. The latter arises from odd numbers of internal reflections of the pump. 
Analogous to the time-domain description of Eqn.\ \ref{eqNLP}, we can write the following expression in the frequency domain and solve the integral
\begin{widetext}
\begin{align}
\begin{split}
P^{(NL)}_i(\omega,t_{pr},z) = &\epsilon_0 \chi^{(3)}_{ijkl}\int\int\int^\infty_{-\infty} [E^{(1)}_j(\omega')E^{(2)*}_k(\omega'')\delta(\omega - \omega' + \omega'' - \omega''')  \\
&\;\;\;\;\;\;\;\;\;\;\;\;\;\;\;\;\;\;\;\;\;\;\;\;\;\;\;+ E^{(1)*}_j(\omega')E^{(2)}_k(\omega'') \delta(\omega + \omega' - \omega'' - \omega''')] E^{(3)}_l(\omega''',t_{pr})  \mathrm{d}\omega'\mathrm{d}\omega''\mathrm{d}\omega''', \\
= &\frac{2 \pi \epsilon_0 \sigma^2}{\sqrt{3}}\chi^{(3)}_{ijkl} A^{(1)}_j A^{(2)}_k A^{(3)}_l \exp\left(-\sigma^2 t_{pr}^2/3 - \left(\omega \mp \omega_1 \pm \omega_2 - \omega_3\right)^2/(6\sigma^2)\right) \\
&\times \exp\left(\left[ n^{(1)}_{\mathrm{g},j} n^{(2)}_{\mathrm{g},k} + \mathrm{q} n^{(1)}_{\mathrm{g},j}  n^{(3)}_{\mathrm{g},l} + \mathrm{q} n^{(2)}_{\mathrm{g},k} n^{(3)}_{\mathrm{g},l} - n^{(1)2}_{\mathrm{g},j} - n^{(2)2}_{\mathrm{g},k} - n^{(3)2}_{\mathrm{g},l} \right] \sigma^2 z^2/(3 c_0^2)\right)\\
&\times\exp\left(\left[2 n^{(3)}_{\mathrm{g},l}- \mathrm{q} n^{(1)}_{\mathrm{g},j} -\mathrm{q} n^{(2)}_{\mathrm{g},k} \right] \sigma^2  t_{pr} z/(3c_0)\right)\\
&\times \exp\left(-\mathrm{i}\left[\tfrac{1}{3}\left(\omega \mp \omega_1 \pm \omega_2 + 2\omega_3\right)t_{pr} - \tfrac{1}{c_0}\left( \pm \mathrm{q} \omega_1 n^{(1)}_{j} \mp  \mathrm{q} \omega_2 n^{(2)}_{k}  +	 \omega_3 n^{(3)}_{l}\right)z\right]\right) \\
   &\times\exp\left(\mathrm{i}\left[\mathrm{q} n^{(1)}_{\mathrm{g},j} + \mathrm{q} n^{(2)}_{\mathrm{g},k} + n^{(3)}_{\mathrm{g},l}\right] 
	\left(\omega \mp \omega_1 \pm \omega_2 - \omega_3\right) z/(3c_0)\right),
\label{app:nl_pol}
\end{split}
\end{align}
\end{widetext}
where the different signs before $\omega_1$ and $\omega_2$ originate in the two terms inside the integral. The solution of the integral is given by the sum of the two sign combinations. Note that the effects of field conjugation and propagation inversion $\mathrm{q}$ differ in the condition of energy conservation in $\delta$. The first 3 lines of the solution represent an envelope function while the lowest 2 lines are oscillatory components in $t_{pr}$ and $z$. 
Again, we approximated the symmetry allowed nonlinear tensor elements as constant.  

To retrieve the emitted nonlinear field $E^{(4)}$ we can again make use of the slowly varying envelope approximation in Eqn.\ \ref{eqnlfield4} and \ref{eqsvaa}, where the wave vector $k^{(4)}_i$ can be approximated according to Eqn.\ \ref{app:wave_vector}. 
The spatial integral in Eqn.\ \ref{eqFT} for the emitted nonlinear field now has an analytic solution. The lengthy result is omitted here but it can be sorted according to the two terms in Eqn.\ \ref{app:nl_pol} that lead to two distinct oscillatory components
\begin{align}
\begin{split}
A^{(4,\mathrm{t})}_{\mathrm{q},i}(\omega, t_{pr}, L) =& A^{+}_{\mathrm{q},i}(\omega, t_{pr}, L) e^{-\mathrm{i} t_{pr} \omega^+_{3,\mathrm{q}}(\omega)} \\&+ A^{-}_{\mathrm{q},i}(\omega, t_{pr}, L) e^{-\mathrm{i} t_{pr} \omega^-_{3,\mathrm{q}}(\omega)}.
\end{split}
\label{app:Asol}
\end{align}
Note that $\omega^{\pm}_{3,\mathrm{q}}(\omega)$ correspond to frequencies observed along the experimental parameter $t_{pr}$ that are shifted with respect to the probe frequency at $\omega_3$. 
In a heterodyne detection this nonlinear signal field is combined with the transmitted probe field as described in Eqn.\ \ref{eqsignal}. The latter acts as a local oscillator $A^{(4,\mathrm{t})}_{\mathrm{q},i}A^{(3,\mathrm{t})*}_{j}$ and the product will be modulated at the difference frequency $\omega^\pm_\mathrm{q} = \omega^\pm_{3,\mathrm{q}} -\omega_3$.
The integral over the spectral overlap of the two fields in Eqn.\ \ref{eqsignal} only leads to non-vanishing signal within the probe bandwidth. In this range $\omega^\pm_\mathrm{q}(\omega)$ is an only slowly changing function of $\omega$ and we can obtain a prediction for the observed frequencies by evaluation at the central frequency $\omega_4=\pm(\omega_1 - \omega_2) + \omega_3$
\begin{widetext}
\begin{align}
\begin{split}
\omega^{\pm}_{\mathrm{q}}&(\omega_4)  =  
\left[\pm \mathrm{q}( n^{(1)}_j\omega_1 - n^{(2)}_k\omega_2)
+ n^{(3)}_l\omega_3 - 
      n^{(4)}_i 
          (\omega_1-\omega_2+\omega_3) \right] \frac{\tfrac{\mathrm{q}}{2} (n^{(1)}_{g,j} + n^{(2)}_{g,k}) - n^{(3)}_{g,l}}{n^{(1)2}_{g,j} + n^{(2)2}_{g,k} + n^{(3)2}_{g,l} - n^{(1)}_{g,j} n^{(2)}_{g,k} - \mathrm{q} (n^{(1)}_{g,j} n^{(3)}_{g,l}+  n^{(2)}_{g,k} n^{(3)}_{g,l})}.
\label{app:fappr2}
\end{split}
\end{align}
Eqn.\ \ref{app:fappr2} shows that the Kerr effect in an anisotropic and dispersive material leads to shifts of the initial probe frequency $\omega_3$ by an amount proportional to the phase mismatch $\Delta k_l = \pm (k^{(1)}_j - k^{(2)}_k) + k^{(3)}_l - k^{(4)}_i$ (in square brackets) and a term related to the differences in group propagation. Note that due to anisotropy, the two frequencies arising from the two conjugate field products indicated by $^\pm$ are not degenerate. Furthermore, upon internal reflection of the excitation fields, the pump propagation direction parameter changes from $\mathrm{q}=1$ to $\mathrm{q}=-1$ leading to a total of four distinct frequencies that may be discerned in an experiment. The case where only one of the excitation fields is reflected is also possible but less relevant due to its typically short duration. 
In the experiment described above the central excitation frequencies are identical $\omega_1 = \omega_2$, further simplifying expression \ref{app:fappr2} to 
\begin{align}
\omega^{\pm}_\mathrm{q} = 
\left[\pm \mathrm{q}(n^{(1)}_i - n^{(1)}_j)\omega_1 + (n^{(3)}_i - n^{(3)}_j)\omega_3 \right] 
\frac{\tfrac{\mathrm{q}}{2} (n^{(1)}_{\mathrm{g},i} + n^{(1)}_{\mathrm{g},j}) - n^{(3)}_{\mathrm{g},i}}{n^{(1)2}_{\mathrm{g},i} + n^{(1)2}_{\mathrm{g},j} + n^{(3)2}_{\mathrm{g},i} - n^{(1)}_{\mathrm{g},i} n^{(1)}_{\mathrm{g},j} - \mathrm{q} (n^{(1)}_{\mathrm{g},i} n^{(3)}_{\mathrm{g},i}+ n^{(1)}_{\mathrm{g},j} n^{(3)}_{\mathrm{g},i})}.
\label{app:fappr3}
\end{align}
\end{widetext}
Eqn.\ \ref{app:fappr3} represents a closed expression for the frequencies observed in the anisotropic and dispersive Kerr effect which is the main result of this discussion.
In the absence of birefringence, the differences $n^{(1)}_i - n^{(2)}_j$ and $n^{(3)}_i - n^{(3)}_j$ vanish, thus no oscillations are observed. 
Furthermore, oscillatory solutions require coupling between different crystal axes $\mathbf{e}_{\mathrm{f}}$ and $\mathbf{e}_{\mathrm{s}}$ and are only enabled by the off-diagonal nonlinear tensor elements, here given by $\chi^{(3)}_{\rm{ssff}}$, $\chi^{(3)}_{\rm{sfsf}}$ and $\chi^{(3)}_{\rm{ffss}}$ (note that in this convention the pump fields correspond to the second and third index position). Reducing the solution to coupling via these elements reduces the number of indices in Eqn.\ \ref{app:fappr3} from 4 to 2 in Eqn.\ \ref{app:fappr4} as equivalent permutations are dropped out. 
 It is easily seen that coupling to the other allowed tensor elements $\chi^{(3)}_{\rm{ssss}}$, $\chi^{(3)}_{\rm{ffff}}$ and $\chi^{(3)}_{\rm{sffs}}$ leads to vanishing frequencies $\omega^{\pm}_\mathrm{q}$. For these terms, propagation effects, however, still cause extended temporal envelopes as a result of group walk-off between pump and probe beams \cite{Maehrlein2020}. 

\subsection{Qualitative picture for Kerr effect oscillations}

\begin{figure}[tbp]
		\includegraphics[width=.475\textwidth]{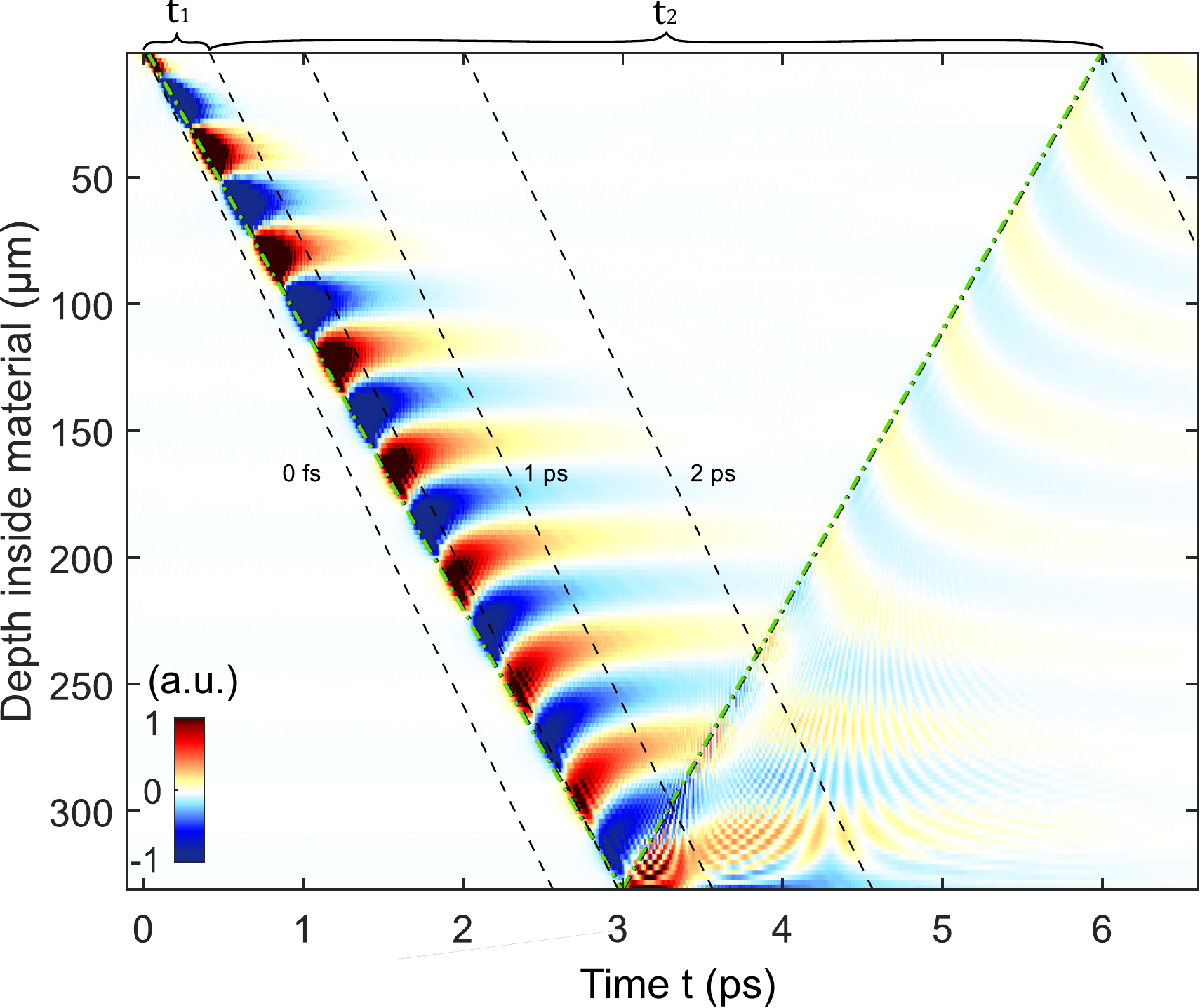}
	\caption{Spatio-temporal representation of $\mathrm{Re}(E^{(1)}_{\rm{s}}E^{(2)*}_{\rm{f}})$ as function of real time $t$ and coordinate $z$ inside a 320 $\mu$m thick sample at $\tau = 0$ fs. Oscillations originate in the anisotropy and the associated phase retardation between fast and slow polarization axes. 
The center of the probe wave-packet is indicated in dashed lines for different pump-probe delays ($t_{pr} = 0, 1, 2,..$ ps) and propagates faster than the fastest spectral component of the pump shown as green dash-dotted line. The pump-probe interaction can be separated in co- and counter-propagation temporal regimes $t_1$ and $t_2$. Due to GVD these regions will differ for different spectral components of the pump.  }
\label{figconstruction}
\end{figure}

An intuitive description of the observed oscillations can be obtained by introducing an additional approximation: In a two-color experiment, the difference in group index anisotropy $n^{(h)}_{\mathrm{g},\mathrm{f}} - n^{(h)}_{\mathrm{g},\mathrm{s}}$ is typically small compared to the dispersive difference $n^{(1)}_{\mathrm{g},i} - n^{(3)}_{\mathrm{g},i}$. Eqn.\ \ref{app:fappr3} is then simplified to
\begin{align}
\begin{split}
\omega^{\pm}_\mathrm{q} &\approx  \frac{\pm \mathrm{q}(n^{(1)}_i - n^{(1)}_j)\omega_1  + (n^{(3)}_i - n^{(3)}_j) \omega_3}{\mathrm{q} n^{(1)}_{\mathrm{g},i} - n^{(3)}_{\mathrm{g},i}}\\
&\approx  \frac{\mp\Delta n^{(1)}\omega_1 - \Delta n^{(3)}\omega_3}{n^{(1)}_{\mathrm{g}}\pm n^{(3)}_{\mathrm{g}}} = \frac{c_0\Delta k^{\pm}}{n^{(1)}_\mathrm{g}\pm n^{(3)}_\mathrm{g}}.
\label{app:fappr4}
\end{split}
\end{align}
The frequency shift in Eqn.\ \ref{app:fappr4} thus depends on the phase mismatch $\Delta k$, which is defined by the birefringence at excitation and probe frequencies $\Delta n^{(h)} = n^{(h)}_\mathrm{f} - n^{(h)}_\mathrm{s}$, as well on the average difference or sum in group indices. The spatio-temporal character of the process can be visualized by a simple geometric construction shown in Fig.\ \ref{figconstruction}. Here, 
anisotropic pump propagation leads to locally varying phases that will be imprinted on the nonlinear polarization when combined with the probe field. Depending on the sample thickness and birefringence, 
the field product $E^{(1)}_{\mathrm{s}}E^{(2)*}_{\mathrm{f}}$ 
undergoes a fixed number of phase oscillations $N^{(1)\pm} = \pm(n^{(1)}_i - n^{(1)}_j) \omega_1 d/(2\pi c_0) $ before arriving at the crystal's backside. 
In addition, for coupling via $\chi^{(3)}_{\mathrm{ssff}}$, where the probe field and nonlinear polarization correspond to index position 4 and 1, respectively, the newly generated nonlinear polarization along $\mathbf{e}_{\mathrm{s}}$ inherits the phase of the probe field polarized along $\mathbf{e}_{\mathrm{f}}$. Birefringence then gives rise to an additional phase-evolution $N^{(3)}= (n_i^{(3)} - n_j^{(3)})\omega_3d/(2\pi c_0)$.
The emitted nonlinear field will thus undergo $N^\pm = \Delta k^\pm d = N^{(1)\pm} + N^{(3)}$ oscillations when the probe delay is scanned. The time over which these oscillations are observed can be found as follows:
In the two-color experiment, the probe pulse propagates faster inside the material allowing for pump-probe overlap inside the sample volume even at significantly positive delays. Different spectral components of the excitation fields arrive at the backside at different times due to group velocity dispersion (GVD) 
as indicated in Fig.\ \ref{figresults} a), c). 
Assuming a probe field polarized along $\mathbf{e}_{\mathrm{f}}$ the latest overlap of three co-propagating fields occurs at the backside at 
\begin{align}
t_1(\omega_1) = [n^{(1)}_{\mathrm{g},\mathrm{f}}(\omega_1) - n^{(3)}_{\mathrm{g},\mathrm{f}}(\omega_3)]d/c_0.
\label{app:t1}
\end{align} 
The observed frequency is thus $\omega^\pm \approx N^\pm/t_1$ as already found in Eqn.\ \ref{app:fappr4} by the difference term in the denominator.
Overlap of the $\mathbf{e}_{\mathrm{s}}$-polarized forward propagating excitation field and the probe field lasts slightly longer until $t'_1 = [n^{(1)}_{\mathrm{g},\mathrm{s}} - n^{(3)}_{\mathrm{g},\mathrm{f}}]d/c_0$. At times $t_1 < t < t'_1$ these two fields can, however, mix with previously reflected components of the $\mathbf{e}_\mathrm{f}$-polarized excitation fields leading to a phase and amplitude discontinuity of the nonlinear signal. 
At $t > t'_1$ spatio-temporal overlap of the three fields is only possible with two reflected excitation fields. 
Due to the inversion of propagation direction, overlap is now possible over a longer temporal interval 
\begin{align}
t_2(\omega_1) = [n^{(1)}_{\mathrm{g},\mathrm{f}}(\omega_1) + n^{(3)}_{\mathrm{g},\mathrm{f}}(\omega_3)]d/c_0
\label{app:t2}
\end{align}
as shown in Fig.\ \ref{figconstruction}. The frequency found in this temporal domain is then simply given by the sum in the denominator in Eqn.\ \ref{app:fappr4}.
After $t_1 + t_2$, the second internal reflection occurs and the process continues at a much reduced intensity.
The timescales of the internal reflections defined by $t_1$ and $t_2$ also govern the non-oscillatory envelope functions in Eqn.\ \ref{app:Asol} and thus define the temporal response also in isotropic materials \cite{Maehrlein2020}.  

\subsection{Energy assignment in 2D spectroscopy}
\label{sec:enass}
\begin{figure}[tb]
	\centering
		\includegraphics[width=.485\textwidth]{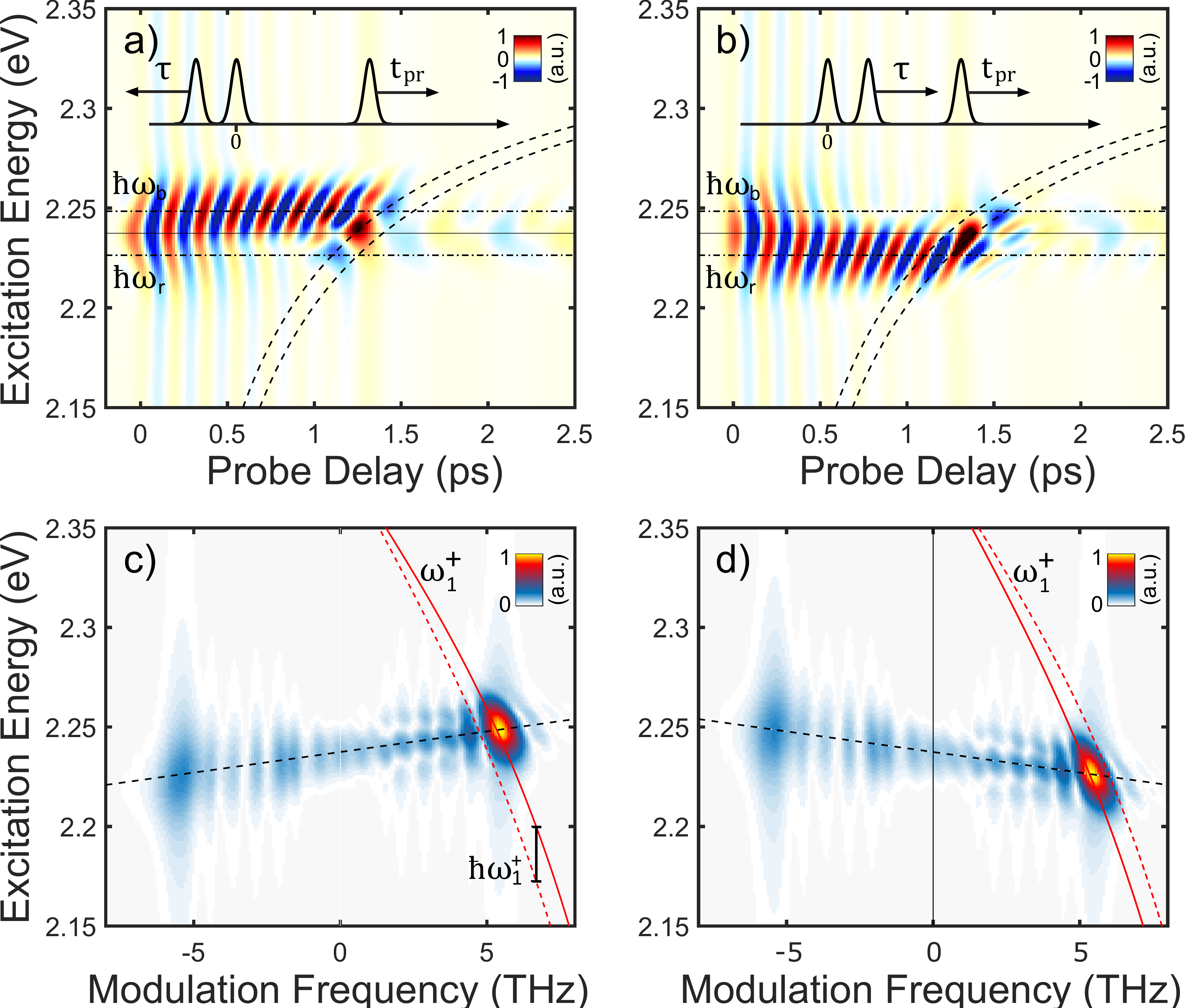}
	\caption{a), b) Simulated Kerr effect with narrow-band excitation of a 240 $\mu$m thick sample for "backward" and "forward" scanning direction of $\tau$, respectively. The central excitation energy $\omega_0$ is shown as solid line at 2.237 eV. The modulation at frequency $\omega^{+}_1$ appears centered on $\omega_b$ or $\omega_r$ in a) and b), respectively. c), d) Corresponding 2D Fourier transformations indicate that the energy assignment is shifted to $E = \hbar \omega_0 \pm \hbar \omega_{m}/2$. Energy assignments of the frequencies $\omega^+_1$ shown as solid lines were corrected accordingly.}
	\label{fig:enass}
\end{figure}

In 2D spectroscopy the resolution in excitation energy is obtained from controlling the relative phase between the fields $E^{(1)}$ and $E^{(2)}$ via the delay $\tau$ and observing the signal dependence in analogy to Fourier-transform infrared spectroscopy \cite{Maehrlein2020, Zanni2009}. However, if the detected signal is a phase dependent oscillatory response, e.g.\ arising from the process described by Eqn.\ \ref{app:fappr3} or due to a coherent Raman mode, corrections to the standard energy assignment are required. The correction can be derived by assuming a two pulse excitation process of an oscillatory response. Here we choose an impulsive stimulated Raman response with negligible damping and frequency $\omega_{m}$ that induces a signal $\propto Q(t,\tau)$
\begin{align}
\begin{split}
E^{(1)}(t) = &E_0 e^{-t^2/(2\sigma_t^2)}e^{-\mathrm{i} \omega_0 t},\\
E^{(2)}(t,\tau) = &E_0 e^{-(t\mp\tau)^2/(2\sigma_t^2)}e^{-\mathrm{i} \omega_0 (t\mp\tau)},\\
R(t) = 
&\begin{cases}
    R_0\sin{(\omega_{m}t)},&  t\geq 0\\
    0,              & t < 0.
\end{cases}\\
Q(t,\tau) = &\int^\infty_{-\infty}\Bigl(E^{(1)}(t')E^{(2)*}(t',\tau)\\
&+ E^{(1)*}(t')E^{(2)}(t',\tau)\Bigr) R(t-t') \mathrm{d}t',
\label{eq:enassignment}
\end{split}
\end{align} 
where the two excitation fields have a center frequency of $\omega_0$, $R_0$ is the Raman coefficient and $\sigma_t$ defines the pulse length. In a 2D time-domain experiment, the choice of scanning $\tau$, here defined as absolute delay, from $\tau = 0$ in the direction of $t$ ("forward", $-$) or against it ("backward", $+$) sets the sign in Eqn.\ \ref{eq:enassignment}. 
For $t - \tau/2> \sigma_t$, the Raman coordinate can be approximated as 
\begin{align}
\begin{split}
Q(t,\tau) &\approx R_0 E_0^2\sqrt{\pi}\sigma_t e^{-(\tau^2/\sigma_t^2 + \sigma_t^2 \omega_{m}^2)/4}\\
&\times\left[\sin(\tau [\omega_0 \pm \frac{\omega_m}{2}] - t\omega_m) 
- \sin(\tau [\omega_0 \mp \frac{\omega_m}{2}] + t \omega_m)\right].
\label{eq:enassignment2}
\end{split}
\end{align}
According to Eqn.\ \ref{eq:enassignment2}, the signal $Q(t,\tau)$ appears as two distinct sidebands $\omega_0 \pm \omega_m/2$ and $\omega_0 \mp \omega_m/2$ in the excitation parameter $\tau$. Furthermore, these two terms are associated with the modulation frequencies $-\omega_m$ and $+\omega_m$ in the frequency domain that corresponds to time $t$. 
Adding dispersion and anisotropy to this process can break the symmetry in amplitude between the two terms in Eqn.\ \ref{eq:enassignment2} and favor one of them. \\
While this energy shift was derived for a Raman-type process, it is also valid for the birefringence induced oscillatory response in the non-resonant case described by Eqn.\ \ref{app:fappr3}. This can be seen from the narrow-band excitation simulations shown in Fig.\ \ref{fig:enass}. 
Thus for "backward" and "forward" scanning of $\tau$, the signals in the positive frequency plane of $t$ are blue- or redshifted in their energy assignment according to 
\begin{align}
E_{b,r} = \hbar \omega_{b,r} =  \hbar \omega_0 \pm \hbar \omega_{m}/2, 
\label{eq:enass}
\end{align}  
where $\omega_m$ now refers to an arbitrary modulation frequency.
This dynamic energy assignment shift needs to be taken into account when comparing 2D experiments and numerical simulations to the analytic expressions in Eqn.\ \ref{app:fappr3} and \ref{app:fappr4}. Accordingly, we find a vertical energy difference along the excitation energy axis of $\hbar \omega^{\pm}_\mathrm{q}$ between "forward" and "backward" scanning direction in experiments as well as simulations.  
 
Note that, this behavior is not commonly observed in conventional 2D spectroscopy in the visible as population dynamics are rarely oscillatory in nature and the relevant resolution in excitation is typically significantly coarser \cite{Nemeth2010}. %

\subsection{Application for birefringence spectroscopy}

The strong sensitivity of $\omega^\pm_q$ to birefringence and group index dispersion encourages a spectroscopic application of this nonlinear effect.
If all four branches are observed over an extended spectral range, the four equations in Eqn.\ \ref{app:fappr4} may allow to solve for the birefringence at excitation and probe frequencies $\Delta n^{(1)}$, $\Delta n^{(3)}$ and the group indices $n^{(1)}_\mathrm{g}$ and $n^{(3)}_\mathrm{g}$. 
In addition, the experimental data also contains direct information on the group indices if the internal reflection times $t_1$ and $t_2$ are observed. Then Eqn.\ \ref{app:t1} and \ref{app:t2} can be combined to yield direct measurements of $n^{(3)}_\mathrm{g} = (t_2 - t_1) c_0/(2d)$ and $n^{(1)}_\mathrm{g} = (t_1 + t_2) c_0/(2d)$.

However, to incorporate the energy assignment shift in Eqn.\ \ref{eq:enass}, to cope with incomplete experimental information or if the approximation in Eqn.\ \ref{app:fappr4} is not valid, a self-consistent birefringent refractive index model can instead be simultaneously fitted to the experimentally observable frequency branches using Eqn.\ \ref{app:fappr3} and to the available information on reflection times in Eqn.\ \ref{app:t1} and \ref{app:t2}. 
The approximation leading to Eqn.\ \ref{app:fappr3} requires negligible GVD over the mixing bandwidth $(\hbar \omega_r, \hbar \omega_b)$. If this is not the case, experimental data may still be analyzed by fitting the refractive index using the numerical calculation described above. Minor deviations between the analytic functions and numerical results in Fig.\ \ref{figresults} d) may be due to GVD. 

Note that the set of equations is invariant to the absolute value of the refractive index. Nevertheless, by combining the here described Kerr effect propagation spectroscopy technique with a single point measurement of the refractive index along one of the polarization axes (e.g.\ at the probe frequency), complete information on the birefringent refractive index can be extracted over the full excitation spectrum.  

In fact, the refractive index shown in Fig.\ \ref{figrefindex} and used for the numerical calculation was obtained by fitting a parameterized refractive index model (see App.\ \ref{app:index}) in this manner to our experimental results while forcing a refractive index at probe frequencies of 1.9 in accordance with previous measurements \cite{Chen2019}. The agreement between numerical results and experiments is thus self-consistent and verifies the described method to experimentally extract birefringence and group index at excitation frequencies.

\section{Conclusion}

This article discussed the fundamental nonlinear optics underlying the ultrafast Kerr effect in anisotropic and dispersive media. It is shown that a non-resonant instantaneous electronic hyperpolarizability can lead to complex extended temporal responses when the sample dimensions are not chosen carefully. 
As we employ methods borrowed from 2D electronic spectroscopy, we may emphasize the conceptional differences in the underlying process. The analysis of electronic spectroscopy is commonly based on the semi-impulsive approximation \cite{Hamm2011} that assumes instantaneous $\delta$-shaped electric fields interacting with the material. Here, we discuss the exact opposite scenario where the material response is assumed instantaneous and the signal dynamics arise from dispersively stretched electric fields.
While group walk-off between excitation and probe fields is already present in isotropic materials, in this work we focus on the particular responses found in anisotropic systems. The nonlinear mixing process then generally comes with a finite phase-mismatch $\Delta k$. Thus, coupling to off-diagonal tensor elements of type $\chi^{(3)}_{\rm{ffss}}$, that are allowed in most point groups, can result in multiple branches of fast oscillations appearing on different timescales in the Kerr signal. 

To minimize the influence of propagation effects on OKE spectroscopy, the experimental geometry can be chosen based on the anticipated reflection times $t_1$ and $t_2$ in Eqn.\ \ref{app:t1} and Eqn.\ \ref{app:t2}. 
If the experiment is intended to resolve material dynamics as fast as $T_{exp} = 100$ fs, we can require $t_1\ll T_{exp}$ or even $t_2\ll T_{exp}$. This effectively sets an upper limit for the sample thickness $d \ll \tau_{exp} c_0 /(n^{(1)}_{\mathrm{g},i} - n^{(3)}_{\mathrm{g},j})$. In the case of room temperature CsPbBr$_3$ with near band-gap excitation this implies $d \ll 10$ $\mu$m. 
The purely photonic propagation response can otherwise bury and distort the dynamics of the nonlinear response function that is commonly the target of ultrafast OKE spectroscopy. Misinterpretation of propagation effects for the nonlinear response function occurred in previous publications where Kerr spectroscopy was applied to bulk perovskites \cite{Zhu1409, Miyata2017}, which are re-evaluated in ref.\ \cite{Maehrlein2020}. 
Even in the presence of a dominant resonant nonlinear response function, propagation still leads to significant distortions, as shown for a Raman-type response in App.\ \ref{app:phonon}.\\ 
In the current publication, we detail how to simulate the Kerr propagation response and derive compact analytic functions that allow to use the Kerr effect for birefringence spectroscopy and to gain quantitative information on the group index.
The described method allows for high resolution measurements on thick samples while being largely insensitive to surface morphology which presents a common limitations of conventional methods such as ellipsometry \cite{Shokhovets2017}.  

\section*{Acknowledgements}
XYZ acknowledges support for the experiments by the Vannevar Bush Faculty Fellowship through Office of Naval Research grant N00014-18-1-2080 and the method development by the US Department of Energy, Office of Energy Science, grant DE-SC0010692. LH acknowledges support from the Swiss National Science Foundation under project ID 187996. SFM was supported by a Feodor Lynen Fellowship of the Alexander von Humboldt Foundation.

\section*{Data availability}
The data that support the findings of this study are available from the corresponding author upon reasonable request.

\appendix
\section{Tensor symmetry and transformation}
\label{apptensor}
For orthorhombic structures, a Cartesian reference frame is defined by the unit vectors aligned with the orthogonal primitive lattice vectors $\mathbf{e}_\mathrm{a}, \mathbf{e}_\mathrm{b}, \mathbf{e}_\mathrm{c}$. The non-vanishing tensor components are then given by
\begin{align}
\begin{split}
&\chi^{(3)}_{\rm{aaaa}},
\chi^{(3)}_{\rm{bbbb}},
\chi^{(3)}_{\rm{cccc}},\\
&\chi^{(3)}_{\rm{aabb}}= \chi^{(3)}_{\rm{abba}}= \chi^{(3)}_{\rm{abab}},
\chi^{(3)}_{\rm{aacc}}= \chi^{(3)}_{\rm{acca}}= \chi^{(3)}_{\rm{acac}},\\
&\chi^{(3)}_{\rm{bbaa}}= \chi^{(3)}_{\rm{baab}}= \chi^{(3)}_{\rm{baba}},
\chi^{(3)}_{\rm{bbcc}}= \chi^{(3)}_{\rm{bccb}}= \chi^{(3)}_{\rm{bcbc}},\\
&\chi^{(3)}_{\rm{ccaa}}= \chi^{(3)}_{\rm{caac}}= \chi^{(3)}_{\rm{caca}},
\chi^{(3)}_{\rm{ccbb}}= \chi^{(3)}_{\rm{cbbc}}= \chi^{(3)}_{\rm{cbcb}}.
\end{split}
\label{eqtcompabc}
\end{align}
For fields propagating parallel to the surface normal, the orthorhombic system and the $f,s$-axes reference frame share the common $\mathbf{e}_\mathrm{b}$-axis, which we tentatively assign to the slow axis as depicted in Fig.\ \ref{figdetection} b). 
Fields polarized along the fast axis then induce a mixture of $a$ and $c$ components. The tensor components can be transformed by rotations around $\mathbf{e}_\mathrm{b} || \mathbf{e}_\mathrm{s}$ with an angle $\varphi = \pi + \tan{(a/c)}$. 
We can write the corresponding basis transform $R_{hh'}$ with the rotation matrix
\begin{align}
\bm{R} &= \begin{pmatrix}
\cos{\varphi} & 0 & \sin{\varphi}  \\
		0			 & 1 & 	0 		\\ 
-\sin{\varphi} & 0 & \cos{\varphi}  \\
\end{pmatrix}.
\label{eq:transformM}
\end{align}
The tensor in the experimental reference frame can be calculated as 
\begin{align}
\chi^{(3)}_{ijkl} = R_{ii'}R_{jj'}R_{kk'}R_{ll'}\chi^{(3)'}_{i'j'k'l'},
\label{eq:transform}
\end{align}
where $i,j,k,l \in [\mathrm{f},\mathrm{s},\mathrm{z}]$ and $i',j',k',l' \in [\mathrm{a},\mathrm{b},\mathrm{c}]$. 
The relevant nonlinear tensor components in the laboratory reference frame are then 
\begin{align}
\begin{split}
\chi^{(3)}_{\mathrm{ssss}} = &\chi^{(3)}_{\mathrm{bbbb}},\\
\chi^{(3)}_{\mathrm{ffff}} = &\chi^{(3)}_{\mathrm{aaaa}} \cos^4\varphi + \chi^{(3)}_{\mathrm{cccc}} \sin^4\varphi \\ &+ \left(3\chi^{(3)}_{\mathrm{aacc}} + 3\chi^{(3)}_{\mathrm{ccaa}}\right)\cos^2{\varphi} \sin^2{\varphi}, \\
\chi^{(3)}_{\mathrm{\mathrm{ssff}}} = &\chi^{(3)}_{\mathrm{sffs}} = \chi^{(3)}_{\mathrm{sfsf}} = \chi^{(3)}_{\mathrm{aabb}} \cos^2{\varphi} + \chi^{(3)}_{\mathrm{ccbb}}\sin^2{\varphi},\\
\chi^{(3)}_{\mathrm{ffss}} = &\chi^{(3)}_{\mathrm{fssf}} = \chi^{(3)}_{\mathrm{fsfs}} = \chi^{(3)}_{\mathrm{bbaa}} \cos^2{\varphi} + \chi^{(3)}_{\mathrm{bbcc}}\sin^2{\varphi}.\\
\end{split}
\label{eqtcompfsz}
\end{align}

\section{Balanced detection}
\label{app:balanced}
The calculation implements the balanced detection as follows: After the sample, the nonlinear field is combined with the probe beam $\mathbf{E}^{(4,\mathrm{t})} + \mathbf{E}^{(3,\mathrm{t})}$. 
In a balanced detection scheme, the combined beams are commonly transmitted through a linear polarizer oriented at 45$^\circ$ with respect to the initial probe polarization in front of the sample, as shown in Fig.\ \ref{figdetection} a). The detected signal is given by the difference in integrated intensity transmitted through the two polarization channels. Balancing the signal in experiments on anisotropic materials requires either to rotate the polarizer or adding a wave retarder upstream of the polarizer. Here, the calculation implements a $\lambda/2$-waveplate (HWP) placed after the sample to reproduce the setup in Ref.\ \cite{Maehrlein2020}.
For the depicted geometry, the combined action of HWP and the polarizer can be represented by a Jones matrix for the transmitted fields through the polarizer channels $m = 1,2$
\begin{align}
\bm{J}_{m} &= \begin{pmatrix}
\cos{(2 \alpha - \phi_{m})} \cos{\phi_{m}}  & \sin{(2 \alpha - \phi_{m})}\cos{\phi_{m}} \\
\cos{(2 \alpha - \phi_{m})} \sin{\phi_{m}}  & \sin{(2 \alpha - \phi_{m})}\sin{\phi_{m}} \\
\end{pmatrix},
\label{eqjones}
\end{align}
with $\phi_{1,2} = \phi \pm \frac{\pi}{4}$ and $\phi$ being the angle between the incoming probe polarization and the material's fast axis $\bm{e}_f$. The HWP balancing angle $\alpha$ is found numerically by the zero-signal condition without excitation 
\begin{align}
S\propto \int {\left(|\bm{J}_1(\alpha) \mathbf{E}^{(3, t)}|^2 - |\bm{J}_2(\alpha) \mathbf{E}^{(3, t)}|^2 \right)d\omega} = 0.
\label{eqbalance}
\end{align}

\section{Refractive index model \& fitting}
\label{app:index}

The used refractive index model function parameterizes scaling and shifting transformations of a rational base function $n_0(\omega)$ in order to fit the two polarization directions $n_\mathrm{f}$ and $n_\mathrm{s}$. To obtain starting values for $n_0$ we first fitted it to published refractive index data of a CsPbBr$_3$ film sample \cite{Schlaus2019}
\begin{align}
n_0(\nu) = \frac{p_1 \nu^2 + p_2 \nu + p_3}{\nu^2 + q_1 \nu + q_2}.
\label{app:fitrat}
\end{align}
With $\omega = 2\pi \nu$ the fixed parameters $p_i$ and $q_i$ are given in Tab.\ \ref{app:tab1}.
\begin{table}[tbh]
\begin{tabular}{ccccccc}
          & $p_1$                       & $p_2$                    & $p_3$   & $q_1$                    & $q_2$ & $q_3$ \\
Unit      & $10^{-24}$Hz$^{-2}$        & $10^{-12}$Hz$^{-1}$     & $10^3$       &$10^{-12}$Hz$^{-1}$      & $10^3$     & $ 10^{12}$Hz \\
\hline
Value     & 2.022                      & -2629                 & 843.4  & -1282                  & 40.64 & 2.0   \\
\end{tabular}
\caption{Fixed parameters of the refractive index model.}
\label{app:tab1}
\end{table}

To fit the birefringent refractive index we parameterized transformations to $n_0(\omega)$ as follows
\begin{align}
\begin{split}
n_{\mathrm{f}}(\nu) &= a_1 \left(n_0(\nu + a_2) + \frac{a_3}{(\nu + a_2)^4 - a_4^4}\right) + a_5 \\
n_{\mathrm{s}}(\nu) &= a_6 n_{\mathrm{f}}(\nu + a_7) + a_8 + a_{9} \frac{q_3^2}{\nu^2 - q_3^2}.
\end{split}
\label{app:fitparam}
\end{align}
The quartic term was added to account for the steeper band-gap response in bulk CsPbBr$_3$, while a 2.0 THz oscillator term in $n_s$ was added to account for lattice induced differences in the low frequency birefringence.
Based on this refractive index model we evaluated Eqn.\ \ref{app:fappr3} for the different branches as well as Eqn.\ \ref{app:t1} for the first reflection timing and fitted parameters $a_i$ to the experimental observations using a genetic algorithm. The global cost function uses equal weights for picosecond differences in $t_1$ and THz differences in $\omega^{\pm}_q$, while $n_\mathrm{f}(800 nm) = 1.9$ is enforced with a large weight. 
To reduce the number of parameters, we set the parameters for spectral translation $a_2$ and splitting $a_7$ to zero. The fit results are given in Tab.\ \ref{app:tab2}.
\begin{table}[h]
\begin{tabular}{cccccccc}
          & $a_1$          & $a_3$    & $a_4$     & $a_5$   & $a_6$    & $a_8$       & $a_9$     \\
Unit      &   1            &  $10^{56}$ Hz$^4$    &  $10^{12}$ Hz   &   1     &     1     &      1       &   1       \\
\hline
Value     & 0.6158        & -35.00  &  580.9     &  0.560   &  1.101  &  -0.1773  & 146.8          
    \\
\end{tabular}
\caption{Fitted parameters of the refractive index model.}
\label{app:tab2}
\end{table}

\section{Phonon line shape with anisotropy \& dispersion}
\label{app:phonon}
\begin{figure}
	\centering
		\includegraphics[width=.485\textwidth]{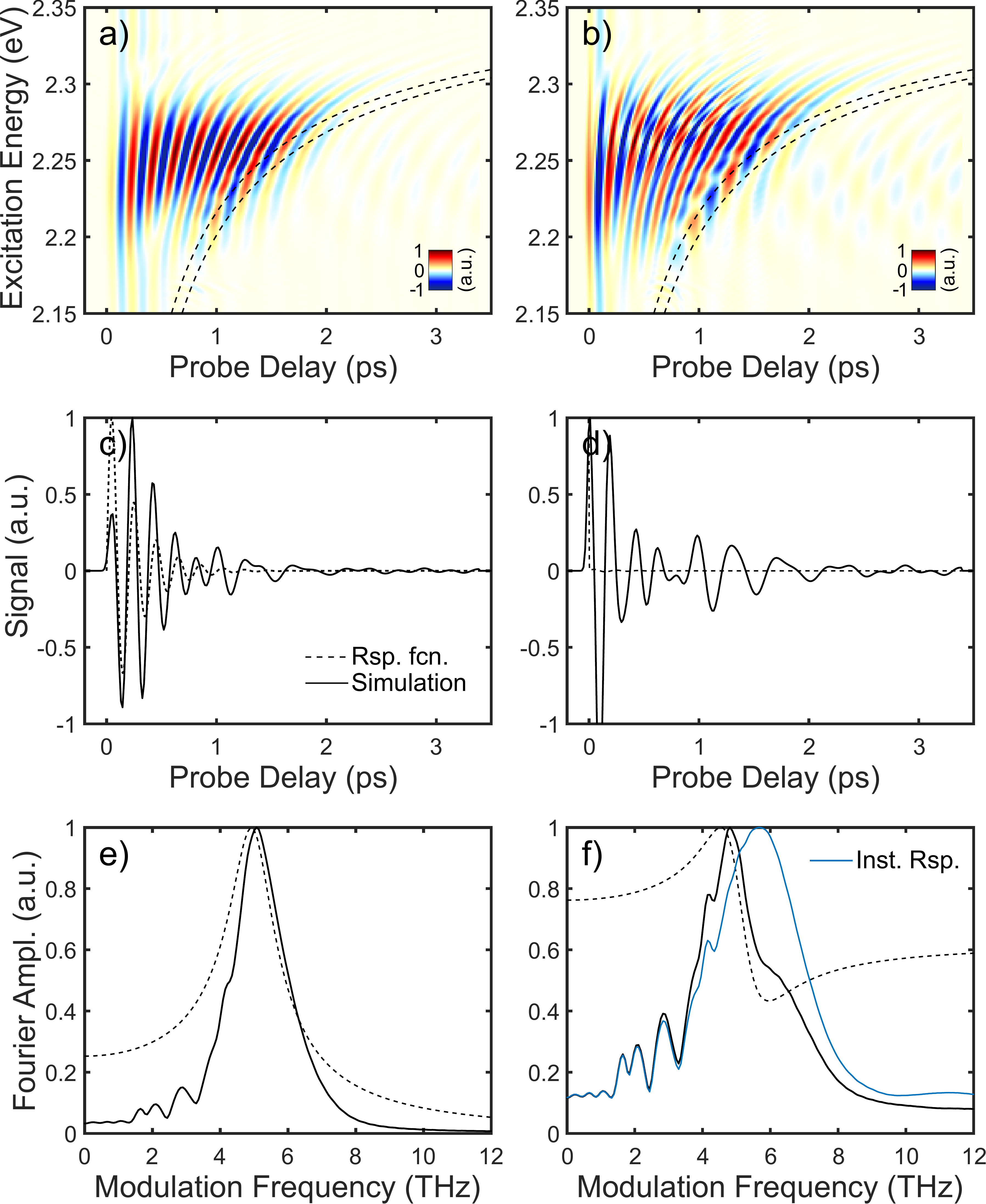}
	\caption{a), b) Simulated time-resolved spectra for a 240 $\mu$m thick sample based on a) a purely phononic response function and b) a combined instantaneous and phononic response function. The corresponding pump-probe traces are shown as solid lines in c) and d), respectively. The employed response functions are shown in broken lines. Their Fourier transforms are shown in e) and f) indicating that dispersion can lead to a narrowing, shift and distortion of a phonon-signature. For comparison, the simulated spectrum with a purely instantaneous response is shown as blue solid line in f).}
	\label{fig:phonon}
\end{figure}

In case the material response function contains Raman resonances, the presence of anisotropy and dispersion gives rise to distortions and shifts of the detected signal with respect to the intrinsic Raman spectrum. As an example, we present calculation results for an orthorhombic material with a $\omega_{ph}= 5.0\times 2\pi$ THz Raman response in all symmetry allowed tensor components. The same refractive index as for CsPbBr$_3$ is used (see App.\ \ref{app:index}).
The phonon response function may be given as
\begin{align}
\begin{split}
\chi^{(3)}_{\mathrm{R}}(t, t', t'', t''') = &\chi^{(3)}_{\mathrm{R},0}\delta(t'-t'')\delta(t-t''')e^{-\Gamma(t-t')} \\
&\times\sin(\sqrt{\omega_{ph}^2 - \Gamma^2}(t-t')),
\label{eq:respphonon}
\end{split}
\end{align}
where $\Gamma=$ 4.0 THz is the phonon damping coefficient. 
Instead of using expression \ref{eqnonlinearpol}, the nonlinear polarization in Eqn.\ \ref{eqNLP} is then obtained from a convolution integral. We calculated the detected signal from a 240 $\mu$m thick sample for a purely phononic $\chi^{(3)}_{\mathrm{R}}$ and a combined phononic and instantaneous response $\chi^{(3)}_{\mathrm{R}} + \chi^{(3)}_{\mathrm{NR}}$. For the latter, the integrated spectral amplitude of the response in the range $[0,10]$ THz was chosen such that the instantaneous contribution is 3-times larger than the phononic one. The results are shown in Fig.\ \ref{fig:phonon}. The same optical parameters as for the calculation shown in Fig.\ \ref{figresults} were used.

\input{adk_lh_bckp.bbl}

\end{document}

%% file: adk_lh_bckp.bbl
%